\documentclass[sigconf]{acmart}

\AtBeginDocument{%
  \providecommand\BibTeX{{%
    \normalfont B\kern-0.5em{\scshape i\kern-0.25em b}\kern-0.8em\TeX}}}

\usepackage{booktabs}
\usepackage{array}
\usepackage{balance} 
\usepackage{multirow}
\usepackage[normalem]{ulem}
\usepackage{color}
\definecolor{lightgray}{RGB}{215,215,215}
\usepackage{colortbl}  
\usepackage{xcolor}
\useunder{\uline}{\ul}{}
\usepackage{subfigure}
\usepackage{algorithm}  
\usepackage{algorithmicx}  
\usepackage[noend]{algpseudocode}  

\usepackage{amsmath}  
\usepackage{enumitem}
\usepackage{tabularx}
\usepackage[utf8]{inputenc}
\usepackage[english]{babel}
\usepackage{amsthm}
\usepackage{bm}
\usepackage{threeparttable}
\newcommand{\ie}{\emph{i.e., }}
\newcommand{\eg}{\emph{e.g., }}

\newcommand{\wrt}{\emph{w.r.t. }}
\newcommand{\cf}{\emph{cf. }}

\newcommand{\imp}[1]{\textcolor{red}{#1}} 

\newlength\myindent
\floatname{algorithm}{Algorithm}

\copyrightyear{2023}
\acmYear{2023}
\setcopyright{acmlicensed}\acmConference[SIGIR '23]{Proceedings of the 46th International ACM SIGIR Conference on Research and Development in Information Retrieval}{July 23--27, 2023}{Taipei, Taiwan}
\acmBooktitle{Proceedings of the 46th International ACM SIGIR Conference on Research and Development in Information Retrieval (SIGIR '23), July 23--27, 2023, Taipei, Taiwan}
\acmPrice{15.00}
\acmDOI{10.1145/3539618.3591663}
\acmISBN{978-1-4503-9408-6/23/07}

\begin{document}


\title{Diffusion Recommender Model}

\author{Wenjie Wang}
\email{wenjiewang96@gmail.com}
\affiliation{
\institution{National University of Singapore}
\country{}
}

\author{Yiyan Xu}
\email{yiyanxu24@gmail.com}
\affiliation{
\institution{University of Science and Technology of China}
\country{}
}

\author{Fuli Feng}
\authornote{Corresponding author: Fuli Feng. This research is supported by the National Key Research and Development Program of China (2020YFB1406703), the National Natural Science Foundation of China (62272437), and Huawei International Pte Ltd.}
\email{fulifeng93@gmail.com}
\affiliation{
\institution{University of Science and Technology of China}
\country{}
}

\author{Xinyu Lin}
\email{xylin1028@gmail.com}
\affiliation{
\institution{National University of Singapore}
\country{}
}

\author{Xiangnan He}
\email{xiangnanhe@gmail.com}
\affiliation{
\institution{University of Science and Technology of China}
\country{}
}

\author{Tat-Seng Chua}
\email{dcscts@nus.edu.sg}
\affiliation{
\institution{National University of Singapore}
\country{}
}

\renewcommand{\shortauthors}{Wenjie Wang et al.}

\begin{abstract}
Generative models such as Generative Adversarial Networks (GANs) and Variational Auto-Encoders (VAEs) are widely utilized to model the generative process of user interactions. However, they suffer from intrinsic limitations such as the instability of GANs and the restricted representation ability of VAEs. Such limitations hinder the accurate modeling of the complex user interaction generation procedure, such as noisy interactions caused by various interference factors. In light of the impressive advantages of \textit{Diffusion Models} (DMs) over traditional generative models in image synthesis, we propose a novel \textit{\textbf{Diff}usion \textbf{Rec}ommender Model} (named DiffRec) to learn the generative process in a denoising manner. To retain personalized information in user interactions, DiffRec reduces the added noises and avoids corrupting users' interactions into pure noises like in image synthesis. In addition, we extend traditional DMs to tackle the unique challenges in recommendation: high resource costs for large-scale item prediction and temporal shifts of user preference. To this end, we propose two extensions of DiffRec: L-DiffRec clusters items for dimension compression and conducts the diffusion processes in the latent space; and T-DiffRec reweights user interactions based on the interaction timestamps to encode temporal information. We conduct extensive experiments on three datasets under multiple settings (\eg clean training, noisy training, and temporal training). The empirical results validate the superiority of DiffRec with two extensions over competitive baselines. 

\vspace{-0.3cm}
\end{abstract}


\maketitle

\vspace{-0.3cm}
\section{Introduction}
\label{sec:introduction}

Generative models such as Generative Adversarial Networks (GANs) and Variational Auto-Encoders (VAEs) have been broadly utilized for personalized recommendation~\cite{liang2018variational, yu2019vaegan, wang2017irgan}. 
Generally speaking, generative recommender models learn the generative process to infer the user interaction probabilities over all non-interacted items. 
Such generative process typically assumes that
users' interaction behaviors with items (\eg clicks) are determined by some latent factors (\eg user preference). 
Due to aligning with the real-world interaction generation procedure, generative recommender models have achieved significant success~\cite{liang2018variational, wang2017irgan}.

\begin{figure}[t]
\setlength{\abovecaptionskip}{0cm}
\setlength{\belowcaptionskip}{-0.5cm}
\centering
\includegraphics[scale=0.83]{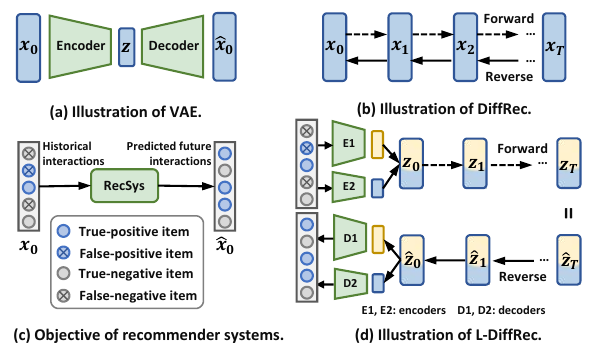}
\caption{Illustration of VAE, DiffRec, the objective of recommender systems, and L-DiffRec.}
\label{fig:intro_example}
\end{figure}

Generative recommender models mainly fall into two groups:
\vspace{-0.1cm}
\begin{itemize}[leftmargin=*]
    \item GAN-based models utilize a generator to estimate users' interaction probabilities and leverage adversarial training to optimize the parameters~\cite{wang2017irgan, jin2020sampling}. 
    However, adversarial training is typically unstable, leading to unsatisfactory performance. 
    
    \item VAEs-based models use an encoder to approximate the posterior distribution over latent factors and maximize the likelihood of observed interactions (Figure~\ref{fig:intro_example}(a))~\cite{liang2018variational, ma2019learning}. 
    While VAEs typically outperform GANs in recommendation, VAEs suffer
    from the trade-off between tractability and representation ability~\cite{sohl2015deep, kingma2016improved}. Tractable and simple encoders might not well capture heterogeneous user preference while the posterior distribution of complex models is likely to be intractable~\cite{sohl2015deep}.
\end{itemize}

\textit{Diffusion Models} (DMs)~\cite{ho2020denoising,sohl2015deep} have achieved state-of-the-art results in image synthesis tasks~\cite{rombach2022high}, which alleviate the trade-off by gradually corrupting the images in a tractable forward process and learning the reverse reconstruction iteratively. 
As shown in Figure~\ref{fig:intro_example}(b), DMs forwardly corrupt $\bm{x}_0$ with random noises step by step, and recover $\bm{x}_0$ from corrupted $\bm{x}_T$ iteratively. 
This forward process leads to a tractable posterior~\cite{sohl2015deep}, and also opens the door to iteratively modeling complex distributions by flexible neural networks in the reverse generation. 
The objectives of recommender models align well with DMs since recommender models essentially infer the future interaction probabilities based on corrupted historical interactions (Figure~\ref{fig:intro_example}(c)), where corruption implies that the interactions are noisy due to false-positive and false-negative items~\cite{sato2020unbiased, wang2021denoising}. 
As such, exploring DMs for recommendation has great potential to
model the complex interaction generation more accurately with strong representation ability.

We propose a \textit{\textbf{Diff}usion \textbf{Rec}ommender Model} named DiffRec, which infers users' interaction probabilities in a denoising manner. Technically, DiffRec gradually corrupts users' interaction histories by injecting scheduled Gaussian noises in the forward process, and then recovers original interactions from the corrupted interactions iteratively via a parameterized neural network. 
Nevertheless, we cannot directly graft the forward process in the image domain due to the necessity of generating personalized recommendations.
To retain personalized information in users' corrupted interactions, we should avoid corrupting users' interaction histories into pure noises like in image synthesis. 
We thus significantly decrease the added noise scales in the forward process (see Section~\ref{sec:method_discussion}).

Taking one step further, we handle two essential challenges in building generative models for recommendation: large-scale item prediction and temporal modeling. 
In detail, 1) generative models require extensive resource costs as predicting the interaction probabilities of all items simultaneously~\cite{liang2018variational}, limiting their application to large-scale item recommendation;
and 2) generative models have to capture the temporal information in the interaction sequence, which is crucial for handling user preference shifts~\cite{xie2021adversarial}. 
To this end, we further extend DiffRec to \textit{Latent DiffRec} (named L-DiffRec) and \textit{Temporal DiffRec} (named T-DiffRec).
\begin{itemize}[leftmargin=*]
    \item L-DiffRec clusters items into groups, compresses the interaction vector over each group into a low-dimensional latent vector via a group-specific VAE, and conducts the forward and reverse diffusion processes
    in the latent space (Figure~\ref{fig:intro_example}(d)). 
    Owing to the clustering and latent diffusion, L-DiffRec significantly reduces the model parameters and 
    memory costs, enhancing the ability of large-scale item prediction (see Section~\ref{sec:l-diffrec} and \ref{sec:performance_l-diffrec}). 
    
    \item T-DiffRec models the interaction sequence via a simple yet effective time-aware reweighting strategy. Intuitively, users' later interactions are assigned with larger weights, and then fed into DiffRec for training and inference (see Section~\ref{sec:t-diffrec} and \ref{sec:performance_t-diffrec}). 
\end{itemize}
We conduct extensive experiments on three representative datasets and compare DiffRec with various baselines under multiple settings (\eg clean training, noisy training with natural or random noises, and temporal training), validating the superiority of our proposed DiffRec and two extensions. We release our code and data at \url{https://github.com/YiyanXu/DiffRec}.

To sum up, the contributions of this work are as follows.
\begin{itemize}[leftmargin=*]
    \item We propose a novel Diffusion Recommender Model, a totally new recommender paradigm that points out a promising future direction for generative recommender models. 

    \item We extend conventional Diffusion Models to reduce the resource costs for high-dimensional categorical predictions and enable the time-sensitive modeling of interaction sequences.
    
    \item We conduct substantial experiments on three datasets under various settings, demonstrating remarkable improvements of DiffRec with two extensions over the baselines. 
    
\end{itemize}
\section{Preliminary}
\label{sec:preliminaries}
DMs have achieved impressive success in various fields, 
mainly consisting of forward and reverse processes~\cite{sohl2015deep, ho2020denoising}. 

\vspace{3pt}
\noindent\textbf{$\bullet $ Forward process.}
Given an input data sample $\bm{x}_0\sim q(\bm{x}_0)$, the forward process constructs the latent variables $\bm{x}_{1:T}$ in a Markov chain by gradually adding Gaussian noises in $T$ steps. 
Specifically, DMs define the forward transition $\bm{x}_{t-1}\to \bm{x}_t$ as $q(\bm{x}_t| \bm{x}_{t-1})=\mathcal{N}(\bm{x}_t;\sqrt{1-\beta_t}\bm{x}_{t-1}, \beta_t\bm{I})$, where $t\in\{1,\dots,T\}$ refers to the diffusion step, $\mathcal{N}$ denotes the Gaussian distribution, and $\beta_t\in (0,1)$ controls the noise scales added at the step $t$. If $T\to\infty$, $\bm{x}_T$ approaches a standard Gaussian distribution~\cite{ho2020denoising}.

\vspace{3pt}
\noindent\textbf{$\bullet $ Reverse process.}
DMs learn to remove the added noises from $\bm{x}_{t}$ to recover $\bm{x}_{t-1}$ in the reverse step, aiming to capture minor changes in the complex generation process. Formally, taking $\bm{x}_T$ as the initial state, DMs learn the denoising process $\bm{x}_t\to\bm{x}_{t-1}$ iteratively by $p_\theta(\bm{x}_{t-1}|\bm{x}_t)=\mathcal{N}(\bm{x}_{t-1};\bm{\mu}_\theta(\bm{x}_t,t),\bm{\Sigma}_\theta(\bm{x}_t,t))$, where $\bm{\mu}_\theta(\bm{x}_t,t)$ and $\bm{\Sigma}_\theta(\bm{x}_t,t)$ are the mean and covariance of the Gaussian distribution predicted by a neural network with parameters $\theta$.

\vspace{3pt}
\noindent\textbf{$\bullet $ Optimization.}
DMs are optimized by maximizing the Evidence Lower Bound~(ELBO) of the likelihood of observed input data $\bm{x}_{0}$:
\begin{equation}\label{eq:DM-ELBO}
\small
\begin{aligned}
\log p(\bm{x}_0) & =\log\int p(\bm{x}_{0:T})\mathrm{d}\bm{x}_{1:T} \\
& =\log\mathbb{E}_{q(\bm{x}_{1:T}|\bm{x}_0)}\left[\dfrac{p(\bm{x}_{0:T})}{q(\bm{x}_{1:T}|\bm{x}_0)}\right] \\
& \ge \underbrace{\mathbb{E}_{q(\bm{x}_1|\bm{x}_0)}\left[\log p_\theta(\bm{x}_0|\bm{x}_1)\right]}_{\small(\text{reconstruction term})} - \underbrace{D_{\text{KL}}(q(\bm{x}_T|\bm{x}_0)\parallel p(\bm{x}_T))}_{\small(\text{prior matching term})} \\
& - \textstyle\sum_{t=2}^{T}\underbrace{\mathbb{E}_{q(\bm{x}_t|\bm{x}_0)}\left[D_{\text{KL}}(q(\bm{x}_{t-1}|\bm{x}_t,\bm{x}_0)\parallel p_\theta(\bm{x}_{t-1}|\bm{x}_t))\right]}_{\small(\text{denoising matching term})},
\end{aligned} 
\end{equation}
where 1) the reconstruction term denotes the negative reconstruction error over $\bm{x}_0$; 2) the prior matching term is a constant without trainable parameters and thus ignorable in the optimization; and 3) the denoising matching terms regulate $p_\theta(\bm{x}_{t-1}|\bm{x}_t)$ to align with the tractable ground-truth transition step $q(\bm{x}_{t-1}|\bm{x}_t,\bm{x}_0)$~\cite{luo2022understanding}. In this way, $\theta$ is optimized to iteratively recover $\bm{x}_{t-1}$ from $\bm{x}_{t}$. 
According to~\cite{ho2020denoising}, the denoising matching terms can be simplified as $\textstyle\sum_{t=2}^{T}\mathbb{E}_{t,\bm{\epsilon}}\left[||\bm{\epsilon}-\bm{\epsilon}_\theta(\bm{x}_t,t)||^2_2\right]$, 
where $\bm{\epsilon}\sim\mathcal{N}(\bm{0},\bm{I})$; and $\bm{\epsilon}_\theta(\bm{x}_t,t)$ is parameterized by a neural network (\eg U-Net~\cite{ho2020denoising}) to predict the noises $\bm{\epsilon}$ that determine $\bm{x}_t$ from $\bm{x}_0$ in the forward process~\cite{luo2022understanding}.

\vspace{3pt}
\noindent\textbf{$\bullet $ Inference.}
After training $\theta$, DMs can draw $\bm{x}_T\sim\mathcal{N}(\bm{0},\bm{I})$ and leverage $p_\theta(\bm{x}_{t-1}|\bm{x}_t)$ to iteratively repeat the generation process $\bm{x}_T\to\bm{x}_{T-1}\to\dots\to\bm{x}_0$.
Besides, prior studies consider adding some conditions to realize the controllable generation~\cite{li2022diffusion, rombach2022high}.

\section{Diffusion Recommender Model}
\label{sec:method}
To take advantage of the strong generation ability of DMs, we propose a novel DiffRec to predict users' future interaction probabilities from corrupted interactions. Given users' historical interactions, DiffRec gradually corrupts them by adding noises in a forward process, and then learns to recover original interactions iteratively. 
By such iterative denoising training, DiffRec can model complex interaction generation procedures and mitigate the effects of noisy interactions. 
Eventually, the recovered interaction probabilities are used to rank and recommend non-interacted items. In addition, we present two extensions of DiffRec for large-scale item prediction and temporal modeling to facilitate the use of DiffRec in practical recommender systems. 

\begin{figure}[t]
\setlength{\abovecaptionskip}{0cm}
\setlength{\belowcaptionskip}{-0.35cm}
\centering
\includegraphics[scale=0.8]{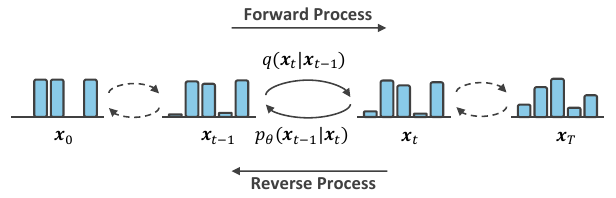}
\caption{An overview of DiffRec, where the histogram denotes the corrupted interactions of a user over all items. The forward process gradually corrupts the user's interaction history by the transition step $q(\bm{x}_t|\bm{x}_{t-1})$, and then the model learns to recover $\bm{x}_0$ using $p_\theta(\bm{x}_{t-1}|\bm{x}_{t})$ step by step.
} 
\label{fig:DifRec}
\end{figure}

\subsection{Forward and Reverse Processes}
As shown in Figure~\ref{fig:DifRec}, DiffRec has two critical processes: 1) a forward process corrupts users' interaction histories by adding Gaussian noises step by step, and 2) a reverse process gradually learns to denoise and output the interaction probabilities.

\vspace{3pt}
\noindent\textbf{$\bullet$ Forward process.}
Given a user $u$ with the interaction history over an item set $\mathcal{I}$, \ie $\bm{x}_u=[x_{u}^{1}, x_{u}^{2},\dots, x_{u}^{|\mathcal{I}|}]$ where $x_{u}^{i}=1$ or $0$ implies whether user $u$ has interacted with item $i$ or not, we can set $\bm{x}_0=\bm{x}_{u}$ as the initial state\footnote{For notation brevity, we omit the subscript $u$ in $\bm{x}_0$ for user $u$.} and parameterize the transition by
\begin{equation}\label{eq:q_xt_given_xt-1}
    q(\bm{x}_t|\bm{x}_{t-1}) = \mathcal{N}(\bm{x}_t; \sqrt{1-\beta_t}\bm{x}_{t-1}, \beta_t\bm{I}),
\end{equation}
where $\beta_t\in(0,1)$ controls the Gaussian noise scales added at each step $t$. 
Thanks to \textit{the reparameterization trick}~\cite{ho2020denoising} and the additivity of two independent Gaussian noises~\cite{luo2022understanding,ho2020denoising}, 
we can directly obtain $\bm{x}_t$ from $\bm{x}_{0}$. Formally,
\begin{equation}
    q(\bm{x}_t|\bm{x}_0) = \mathcal{N}(\bm{x}_t; \sqrt{\bar{\alpha}_t}\bm{x}_0, (1-\bar{\alpha}_t)\bm{I}),
    \label{eq:q_xt_given_x0}
\end{equation}
where $\alpha_t=1-\beta_t$, $\bar{\alpha}_t=\prod_{t'=1}^{t}\alpha_{t'}$, and then we can reparameterize $\bm{x}_t=\sqrt{\bar{\alpha}_t}\bm{x}_0+\sqrt{1-\bar{\alpha}_t}\bm{\epsilon}$ with $\bm{\epsilon}\sim\mathcal{N}(\bm{0},\bm{I})$.
To regulate the added noises in $\bm{x}_{1:T}$, we design a linear noise schedule for $1-\bar{\alpha}_t$, \ie
\begin{equation}
    1 - \bar{\alpha}_t = s\cdot\left[\alpha_{\min} + \dfrac{t-1}{T-1}(\alpha_{\max} - \alpha_{\min})\right],\quad t\in\{1,\dots,T\},
    \label{eq:noise_schedule}
\end{equation}
where a hyper-parameter $s\in\left[0,1\right]$ controls the noise scales, and two hyper-parameters $\alpha_{\min}<\alpha_{\max}\in(0,1)$ indicating the upper and lower bounds of the added noises.

\vspace{3pt}
\noindent\textbf{$\bullet$ Reverse process.}
Starting from $\bm{x}_T$, the reverse process gradually recovers users' interactions by the denoising transition step:
\begin{equation}
    p_\theta(\bm{x}_{t-1}|\bm{x}_t)=\mathcal{N}(\bm{x}_{t-1};\bm{\mu}_\theta(\bm{x}_t,t),\bm{\Sigma}_\theta(\bm{x}_t,t)),
    \label{eq:p_theta}
\end{equation}
where $\bm{\mu}_\theta(\bm{x}_t,t)$ and $\bm{\Sigma}_\theta(\bm{x}_t,t)$ are the Gaussian parameters outputted by any neural networks with learnable parameters $\theta$.

\subsection{DiffRec Training}\label{sec:method_objective}
To learn $\theta$, DiffRec aims to maximize the ELBO of observed user interactions $\bm{x}_0$:
\begin{equation}\label{eq:DiffRec-ELBO}
\small
\begin{aligned}
\log p(\bm{x}_0) & \ge \underbrace{\mathbb{E}_{q(\bm{x}_1|\bm{x}_0)}\left[\log p_\theta(\bm{x}_0|\bm{x}_1)\right]}_{\small(\text{reconstruction term})} \\
& - \textstyle\sum_{t=2}^{T}\underbrace{\mathbb{E}_{q(\bm{x}_t|\bm{x}_0)}\left[D_{\text{KL}}(q(\bm{x}_{t-1}|\bm{x}_t,\bm{x}_0)\parallel p_\theta(\bm{x}_{t-1}|\bm{x}_t))\right]}_{\small(\text{denoising matching term})}.
\end{aligned} 
\end{equation}
Note that the prior matching term in Eq. (\ref{eq:DM-ELBO}) is omitted as it is a constant. 
Besides, the reconstruction term measures the recovery probability of $\bm{x}_0$ while denoising matching terms regulate the recovery of $\bm{x}_{t-1}$ with $t$ varying from $2$ to $T$ in the reverse process. So far, the optimization lies in maximizing the reconstruction term and denoising matching terms.

\vspace{3pt}
\noindent\textbf{$\bullet$ Estimation of denoising matching terms.} The denoising matching term forces $p_\theta(\bm{x}_{t-1}|\bm{x}_t)$ to approximate the tractable distribution $q(\bm{x}_{t-1}|\bm{x}_t,\bm{x}_0)$ via KL divergence. Through 
Bayes rules, $q(\bm{x}_{t-1}|\bm{x}_t,\bm{x}_0)$ can be rewritten as the following closed form~\cite{luo2022understanding}: 
\begin{equation}
\begin{aligned}
&q(\bm{x}_{t-1}|\bm{x}_t,\bm{x}_0) \propto\mathcal{N}(\bm{x}_{t-1};\tilde{\bm{\mu}}(\bm{x}_t, \bm{x}_0, t),\sigma^2(t)\bm{I}), \text{ where} \\ 
\end{aligned}
\end{equation}
\begin{equation}
\left\{
\begin{aligned}
&\tilde{\bm{\mu}}(\bm{x}_t,\bm{x}_0,t) =\dfrac{\sqrt{\alpha_t}(1-\bar{\alpha}_{t-1})}{1-\bar{\alpha}_t}\bm{x}_t+\dfrac{\sqrt{\bar{\alpha}_{t-1}}(1-\alpha_t)}{1-\bar{\alpha}_t}\bm{x}_0, \\
&\sigma^2(t) =\dfrac{(1-\alpha_t)(1-\bar{\alpha}_{t-1})}{1-\bar{\alpha}_t}.
\end{aligned}
\right.
\label{eq:DiffRec_mean}
\end{equation}
$\tilde{\bm{\mu}}(\bm{x}_t,\bm{x}_0,t)$ and $\sigma^2(t)\bm{I}$ are the mean and covariance of $q(\bm{x}_{t-1}|\bm{x}_t,\bm{x}_0)$ derived from Eq. (\ref{eq:q_xt_given_xt-1}) and Eq. (\ref{eq:q_xt_given_x0})~\cite{ho2020denoising}. 
Besides, to keep training stability and simplify the calculation, we ignore the learning of $\bm{\Sigma}_\theta(\bm{x}_t,t)$ in $p_\theta(\bm{x}_{t-1}|\bm{x}_t)$ and directly set $\bm{\Sigma}_\theta(\bm{x}_t,t)=\sigma^2(t)\bm{I}$
by following~\cite{ho2020denoising}. Thereafter, the denoising matching term $\mathcal{L}_t$ at step $t$ can be calculated by
\begin{equation}
    \begin{aligned}
        \mathcal{L}_t & \triangleq \mathbb{E}_{q(\bm{x}_t|\bm{x}_0)}\left[D_{\text{KL}}(q(\bm{x}_{t-1}|\bm{x}_t,\bm{x}_0)\parallel p_\theta(\bm{x}_{t-1}|\bm{x}_t))\right] \\
        & = \mathbb{E}_{q(\bm{x}_t|\bm{x}_0)}\left[\dfrac{1}{2\sigma^2(t)}\left[\parallel\bm{\mu}_\theta(\bm{x}_t,t)-\tilde{\bm{\mu}}(\bm{x}_t,\bm{x}_0,t)\parallel_2^2\right]\right],
    \end{aligned}
   \label{eq:L_t}
\end{equation}
which pushes $\bm{\mu}_\theta(\bm{x}_t,t)$ to be close to $\tilde{\bm{\mu}}(\bm{x}_t,\bm{x}_0,t)$. Following Eq. (\ref{eq:DiffRec_mean}), we can similarly factorize $\bm{\mu}_\theta(\bm{x}_t,t)$ via
\begin{equation}
\label{eq:mu_theta}
    \bm{\mu}_\theta(\bm{x}_t,t) = \dfrac{\sqrt{\alpha_t}(1-\bar{\alpha}_{t-1})}{1-\bar{\alpha}_t}\bm{x}_t+\dfrac{\sqrt{\bar{\alpha}_{t-1}}(1-\alpha_t)}{1-\bar{\alpha}_t}\hat{\bm{x}}_{\theta}(\bm{x}_t,t),
\end{equation}
where $\hat{\bm{x}}_{\theta}(\bm{x}_t,t)$ is the predicted $\bm{x}_0$ based on $\bm{x}_t$ and $t$. Furthermore, by substituting Eq. (\ref{eq:mu_theta}) and Eq. (\ref{eq:DiffRec_mean}) into Eq. (\ref{eq:L_t}), we have
\begin{equation}\label{eq:L_t_x}
\small
    \begin{aligned}
        \mathcal{L}_t = \mathbb{E}_{q(\bm{x}_t|\bm{x}_0)}\left[\dfrac{1}{2}\left(\dfrac{\bar{\alpha}_{t-1}}{1-\bar{\alpha}_{t-1}}-\dfrac{\bar{\alpha}_{t}}{1-\bar{\alpha}_{t}}\right)\parallel\hat{\bm{x}}_{\theta}(\bm{x}_t,t)-\bm{x}_0\parallel_2^2\right],
    \end{aligned}
\end{equation}
which regulates $\hat{\bm{x}}_{\theta}(\bm{x}_t,t)$ to predict $\bm{x}_0$ accurately. 

To summarize, for estimating denoising matching terms, we need to implement $\hat{\bm{x}}_{\theta}(\bm{x}_t,t)$ by neural networks and calculate Eq. (\ref{eq:L_t_x}). Following MultiVAE~\cite{liang2018variational}, we also instantiate $\hat{\bm{x}}_{\theta}(\cdot)$ via a Multi-Layer Perceptron (MLP) that takes $\bm{x}_t$ and the step embedding of $t$ as inputs to predict $\bm{x}_0$. 

\vspace{3pt}
\noindent\textbf{$\bullet$ Estimation of the reconstruction term.}
We define $\mathcal{L}_1$ as the negative of the reconstruction term in Eq. (\ref{eq:DiffRec-ELBO}), and calculate $\mathcal{L}_1$ by 
\begin{equation}
    \begin{aligned}
        \mathcal{L}_1 &\triangleq -\mathbb{E}_{q(\bm{x}_1|\bm{x}_0)}\left[\log p_\theta(\bm{x}_0|\bm{x}_1)\right]\\
        & = \mathbb{E}_{q(\bm{x}_1|\bm{x}_0)}\left[\parallel\hat{\bm{x}}_\theta(\bm{x}_1,1)-\bm{x}_0\parallel_2^2\right],
    \end{aligned}
  \label{eq:L_1}
\end{equation}
where we estimate the Gaussian log-likelihood $\log p(\bm{x}_0|\bm{x}_1)$ by unweighted $-||\hat{\bm{x}}_\theta(\bm{x}_1,1)-\bm{x}_0||_2^2$ as discussed in~\cite{liang2018variational}.

\vspace{3pt}
\noindent\textbf{$\bullet$ Optimization.}
According to Eq. (\ref{eq:L_t_x}) and Eq. (\ref{eq:L_1}), ELBO in Eq. (\ref{eq:DiffRec-ELBO}) 
can be formulated as $-\mathcal{L}_1-{\textstyle\sum_{t=2}^T\mathcal{L}_t}$. Therefore, to maximize the ELBO, we can optimize $\theta$ in $\hat{\bm{x}}_{\theta}(\bm{x}_t,t)$ by minimizing ${\textstyle\sum_{t=1}^T\mathcal{L}_t}$. 
In the practical implementation, we uniformly sample step $t$ to optimize an expectation $\mathcal{L}(\bm{x}_0,\theta)$ over $t\sim \mathcal{U}(1,T)$. Formally,
\begin{equation}\small
    \mathcal{L}(\bm{x}_0,\theta)=\mathbb{E}_{t\sim \mathcal{U}(1,T)}\mathcal{L}_t. 
    \label{eq:L}
\end{equation}
The training procedure of DiffRec is presented in Algorithm~\ref{algo:training}.

\begin{algorithm}[t]
	\caption{\textbf{DiffRec Training}}  
	\label{algo:training}
	\begin{algorithmic}[1]
		\Require all users' interactions $\Bar{\bm{X}}$ and randomly initialized $\theta$.
            \Repeat 
            \State Sample a batch of users' interactions $\bm{X}\subset\Bar{\bm{X}}$.
            \ForAll{$\bm{x}_0\in \bm{X}$}
            \State Sample $t\sim\mathcal{U}(1,T)$ or $t\sim p_t$, $\bm{\epsilon}\sim\mathcal{N}(\bm{0},\bm{I})$;
            \State Compute $\bm{x}_t$ given $\bm{x}_0$, $t$, and $\bm{\epsilon}$ via $q(\bm{x}_t|\bm{x}_0)$ in Eq. (\ref{eq:q_xt_given_x0});
            \State {Compute $\mathcal{L}_t$ by Eq. (\ref{eq:L_t_x}) if $t>1$, otherwise by Eq. (\ref{eq:L_1});}
            \State Take gradient descent step on $\nabla_\theta\mathcal{L}_t$ to optimize $\theta$;
            \EndFor
            \Until{converged}
            \Ensure optimized $\theta$.
	\end{algorithmic}
\end{algorithm}
\setlength{\textfloatsep}{0.28cm}

\vspace{3pt}
\noindent\textbf{$\bullet$ Importance sampling.}
Since the optimization difficulty might vary across different steps. we consider using \textit{importance sampling}~\cite{nichol2021improved} to emphasize the learning over the steps with large loss values of $\mathcal{L}_t$. Formally, we use a new sampling strategy for $t$: 
\begin{equation}\small
\label{eq:L_importance}
\mathcal{L}^\bigtriangleup(\bm{x}_0,\theta)=\mathbb{E}_{t\sim p_t}\left[\dfrac{\mathcal{L}_t}{p_t}\right],
\end{equation}
where $p_t\propto\sqrt{\mathbb{E}\left[\mathcal{L}_t^2\right]}/\sqrt{\sum_{t'=1}^{T}\mathbb{E}\left[\mathcal{L}_{t'}^2\right]}$ denotes the sampling probability and $\sum_{t=1}^{T} p_t=1$.  
We here calculate $\mathbb{E}\left[\mathcal{L}_t^2\right]$ by collecting ten $\mathcal{L}_t$ values during training and taking the average. 
Before acquiring enough $\mathcal{L}_t$, we still adopt the uniform sampling. 
Intuitively, the steps with large $\mathcal{L}_t$ values will be more easily sampled.

\subsection{DiffRec Inference}
In image synthesis tasks, DMs draw random Gaussian noises for reverse generation, possibly guided by the gradients from a pre-trained classifier or other signals such as textual queries. However, corrupting interactions into pure noises will hurt personalized user preference in recommendation (see empirical evidence in Section~\ref{sec:in_depth}). It is also non-trivial to design additional classifiers or guidance signals. As such, we propose a simple inference strategy to align with DiffRec training for interaction prediction. 

Specifically, DiffRec firstly corrupts $\bm{x}_0$ by $\bm{x}_0\to \bm{x}_1 \to \dots \to \bm{x}_{T'}$ for $T'$ steps in the forward process, and then sets $\hat{\bm{x}}_T=\bm{x}_{T'}$ to execute reverse denoising $\hat{\bm{x}}_T\to \hat{\bm{x}}_{T-1} \to \dots \to \hat{\bm{x}}_{0}$ for $T$ steps. The reverse denoising ignores the variance (like in MultiVAE~\cite{liang2018variational}) and utilize $\hat{\bm{x}}_{t-1}=\bm{\mu}_\theta(\hat{\bm{x}}_t,t)$ via Eq. (\ref{eq:mu_theta}) for deterministic inference. 
In particular, in considering 1) the collected user interactions are naturally noisy due to false-positive and false-negative interactions~\cite{wang2022learning, wang2021denoising,wang2021clicks} and 2) retaining personalized information, we reduce the added noises in the forward process by setting $T'<T$. 
Finally, we use $\hat{\bm{x}}_{0}$ for item ranking and recommend top-ranked items. The inference procedure is summarized in Algorithm~\ref{algo:inference}.

\begin{algorithm}[t]
	\caption{\textbf{DiffRec Inference}}  
	\label{algo:inference}
	\begin{algorithmic}[1]
		\Require $\theta$ and the interaction history $\bm{x}_0$ of user $u$.
            \State Sample $\bm{\epsilon}\sim\mathcal{N}(\bm{0},\bm{I})$.
            \State Compute $\bm{x}_{T'}$ given $\bm{x}_0$, $T'$, and $\bm{\epsilon}$ via Eq. (\ref{eq:q_xt_given_x0}), and set $\hat{\bm{x}}_T=\bm{x}_{T'}$.
            \For{$t=T,\dots,1$}
            \State $\hat{\bm{x}}_{t-1}=\bm{\mu}_\theta(\hat{\bm{x}}_t,t)$ calculated from $\hat{\bm{x}}_t$ and $\hat{\bm{x}}_{\theta}(\cdot)$ via Eq. (\ref{eq:mu_theta});
            \EndFor
            \Ensure the interaction probabilities $\hat{\bm{x}}_0$ for user $u$.
	\end{algorithmic}
\end{algorithm}

\subsection{Discussion}\label{sec:method_discussion}
Unlike image synthesis, we highlight two special points of DiffRec. 
\begin{itemize}[leftmargin=*]
    \vspace{-2pt}
    \item \textbf{Personalized recommendation.} 1) During training, we do not corrupt users' interactions into pure noises for retaining some personalized information; that is, the latent variable $\bm{x}_T$ does not approach the standard Gaussian noises that lose extensive personalized characteristics. It is similar to the selection of $\beta$ in MultiVAE to control the strength of the prior constraint, \ie the KL divergence (see Section 2.2.2 in~\cite{liang2018variational}). In practice, We reduce $s$ and $\alpha_{\max}$ in the noise schedule of Eq. (\ref{eq:noise_schedule}) to lessen the noises. 
    And 2) we also decrease the added noises for inference by controlling $T'<T$ by considering the natural noises in user interactions. 

    \item \textbf{$\bm{x}_0$-ELBO.} DiffRec is optimized by predicting $\bm{x}_0$ instead of $\bm{\epsilon}$ like in Section~\ref{sec:preliminaries} because: 1) the key objective of recommendation is to predict $\hat{\bm{x}}_0$ for item ranking, and thus $\bm{x}_0$-ELBO is intuitively more appropriate for our task; and 2) randomly sampled $\bm{\epsilon}\sim\mathcal{N}(\bm{0},\bm{I})$ is unsteady and forcing an MLP to estimate such a $\bm{\epsilon}$ is more challenging (see empirical analysis in Section~\ref{sec:in_depth}).

\end{itemize} 

\vspace{-0.3cm}
\subsection{Latent Diffusion}\label{sec:l-diffrec}

\begin{figure}[t]
\setlength{\abovecaptionskip}{-0.10cm}
\setlength{\belowcaptionskip}{0cm}
\centering
\includegraphics[scale=0.96]{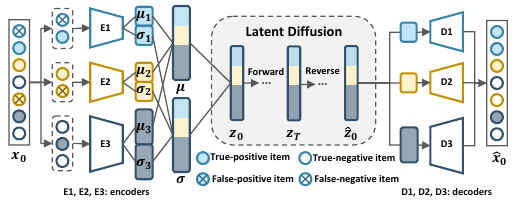}
\caption{Illustration of L-DiffRec. $\bm{z}_0=\bm{\mu}+\bm{\sigma}\odot\bm{\epsilon}$, where $\bm{\epsilon}\sim\mathcal{N}(\bm{0},\bm{I})$. L-DiffRec clusters items for compression via multiple VAEs and conducts latent diffusion.} 
\label{fig:L-DifRec}
\end{figure}

Generative models, such as MultiVAE and DiffRec, predict the interaction probabilities $\hat{\bm{x}}_0$ over all items simultaneously, requiring extensive resources and limiting large-scale item prediction in industry. 
To reduce the costs, we offer L-DiffRec, which clusters items for dimension compression via multiple VAEs and conducts diffusion processes in the latent space as shown in Figure~\ref{fig:L-DifRec}. 

\vspace{3pt}
\noindent\textbf{$\bullet$ Encoding for compression.} 
Given an item set $\mathcal{I}$, L-DiffRec first adopts \textit{k-means} to cluster items into $C$ categories $\{\mathcal{I}_1, \mathcal{I}_2, \dots, \mathcal{I}_C\}$ based on item representations (\eg trained item embeddings from LightGCN). L-DiffRec then divides user interaction vector $\bm{x}_0$ into $C$ parts according to the clusters, \ie $\bm{x}_0\to\{\bm{x}_0^c\}_{c=1}^C$, where $\bm{x}_0^c$ represents the interactions of user $u$ over $\mathcal{I}_c$. 
Afterwards, we use a variational encoder parameterized by $\phi_c$ to compress each $\bm{x}_0^c$ to a low-dimensional vector $\bm{z}_0^c$, where the encoder predicts $\bm{\mu}_{\phi_c}$ and $\sigma^2_{\phi_c}\bm{I}$ as the mean and covariance of the variational distribution $q_{\phi_c}(\bm{z}_0^c|\bm{x}_0^c)=\mathcal{N}(\bm{z}_0^c; \bm{\mu}_{\phi_c}(\bm{x}_0^c),\sigma_{\phi_c}^2(\bm{x}_0^c)\bm{I})$. The clustering can lessen resource costs since it can 1) achieve parallel calculation of different categories and 2) break the full connections among the multiple encoders to save parameters compared to vanilla VAE~\cite{liang2018variational}. 

\vspace{3pt}
\noindent\textbf{$\bullet$ Latent diffusion.} 
By concatenating $\{\bm{z}_0^c\}_{c=1}^C$, we can obtain the compressed $\bm{z}_0$ for diffusion. Like DiffRec training, we replace $\bm{x}_0$ with $\bm{z}_0$ to do the forward and reverse processes in the latent space. Similar to Eq. (\ref{eq:L}), we have the optimization loss as $\mathcal{L}(\bm{z}_0,\theta)=\mathbb{E}_{t\sim \mathcal{U}(1,T)}\mathcal{L}_t$, where $\theta$ marks the parameters of the denoising MLP.

\vspace{3pt}
\noindent\textbf{$\bullet$ Decoding.} 
As shown in Figure~\ref{fig:L-DifRec}, we split the reconstructed $\hat{\bm{z}}_0$ from the reverse process into $\{\hat{\bm{z}}_0^c\}_{c=1}^C$ according to item clusters. 
Each $\hat{\bm{z}}_0^c$ is then fed into a separate decoder parameterized by $\psi_c$ to predict $\hat{\bm{x}}_0$ via $p_{\psi_c}(\hat{\bm{x}}_0^c|\hat{\bm{z}}_0^c)$, which is similar to MultiVAE~\cite{liang2018variational}.

\vspace{3pt}
\noindent\textbf{$\bullet$ Training.} 
Intuitively, the encoder $q_{\phi_c}$ and decoder $p_{\psi_c}$ jointly constitute a VAE that bridges the interaction space and the latent space. Following MultiVAE~\cite{liang2018variational}, the set of VAEs with $\phi=\{\phi_c\}_{c=1}^C$ and $\psi = \{\psi_c\}_{c=1}^C$ could optimized by:
\begin{equation}
\small
\begin{aligned}
\mathcal{L}_v(\bm{x}_0,\phi,\psi) =\sum_{c=1}^C  &[\mathbb{E}_{q_{\phi_c}(\bm{z}_0^c|\bm{x}_0^c)}\left[\log p_{\psi_c}(\bm{x}_0^c|\bm{z}_0^c)\right] \\
& -\gamma \cdot D_{\text{KL}}(q_{\phi_c}(\bm{z}_0^c|\bm{x}_0^c)|| p(\bm{z}_0^c))],
\label{eq:vae-ELBO-lamda}
\end{aligned}
\end{equation}
where $\gamma$ is to control the strength of KL regularization. 
Subsequently, combining the loss of diffusion and VAEs, we have $\mathcal{L}_v(\bm{x}_0,\phi,\psi) + \lambda\cdot\mathcal{L}(\bm{z}_0,\theta)$
for L-DiffRec optimization, where the hyper-parameter $\lambda$ ensures the two terms in the same magnitude. 

\vspace{3pt}
\noindent\textbf{$\bullet$ Inference.} 
For inference, L-DiffRec first splits $\bm{x}_0$ into $\{\bm{x}_0^c\}_{c=1}^C$, and then compresses each $\bm{x}_0^c$ into a deterministic variable $\bm{z}_0^c=\bm{\mu}_{\phi_c}(\bm{x}_0^c)$ without considering variance~\cite{liang2018variational}. After that, L-DiffRec concatenates $\{\bm{z}_0^c\}_{c=1}^C$ into $\bm{z}_0$ for diffusion like DiffRec. Finally, by feeding the reconstructed $\hat{\bm{z}}_0$ into the decoders, we will obtain $\hat{\bm{x}}_0$ for item ranking and generate top-$K$ recommendations.

\subsection{Temporal Diffusion}\label{sec:t-diffrec}
Since user preference might shift over time, it is crucial to capture temporal information during DiffRec learning. Assuming that more recent interactions can better represent users' current preferences, we propose a time-aware reweighting strategy to assign larger weights to users' later interactions.

Formally, for user $u$ with $M$ interacted items, the interaction time is available and the interaction sequence is formulated as $\mathcal{S}=\{i_1, i_2, \dots, i_M\}$, where $i_m$ denotes the ID of the $m$-th interacted item. We define the weights of interacted items $\bm{w} = \left[w_1, w_2, \dots, w_M\right]$ via a time-aware linear schedule\footnote{We use a linear schedule instead of the exponential scaling to simplify the reweighting strategy and save hyper-parameters, leaving more options to future work.}: $w_m = w_{\min} + \dfrac{m-1}{M-1}(w_{\max}-w_{\min})$, 
where the two hyper-parameters $w_{\min}<w_{\max}\in(0,1]$ represent the lower and upper bounds of interaction weights. Thereafter, the interaction history $\bm{x}_0$ of user $u$ is reweighted as $\bar{\bm{x}}_0 = \bm{x}_0 \odot \bar{\bm{w}}$, 
where $\bar{\bm{w}}\in\mathbb{R}^{|\mathcal{I}|}$ is the weight vector calculated by $\bm{w}$, \ie
\begin{equation}
\small
\bar{\bm{w}}[i]=
    \begin{cases}
        \bm{w}[\text{Idx}(i)], & \text{if}~i\in\mathcal{S} \\
        0, & \text{else}
    \end{cases}
\end{equation}
where $\text{Idx}(i)$ denotes the index of item $i$ in the interaction sequence $\mathcal{S}$ of user $u$.
By feeding the reweighted interaction history $\bar{\bm{x}}_0$ into DiffRec and L-DiffRec, we will obtain T-DiffRec and LT-DiffRec using temporal information, respectively.

\section{Experiments}
\label{sec:experiment}
In this section, we conduct extensive experiments on three real-world datasets to answer the following research questions:
\begin{itemize}[leftmargin=*]
    \item \textbf{RQ1:} How does our DiffRec perform compared to the baselines under various experimental settings and how do the designs of DiffRec (\eg importance sampling, the inference step $T'$, and the reduced noise scales) affect the performance?
    \item \textbf{RQ2:} How does L-DiffRec perform regarding the recommendation accuracy and resource costs?
    \item \textbf{RQ3:} Can T-DiffRec surpass sequential recommender models when interaction timestamps are available for training?
    
\end{itemize}

\subsection{Experimental Settings}
\subsubsection{\textbf{Datasets}}\label{sec:datasets}
 \begin{table}[t]
 \setlength{\abovecaptionskip}{0.05cm}
\setlength{\belowcaptionskip}{0.1cm}
\caption{Statistics of three datasets under two different settings, where ``C'' and ``N'' represent clean training and natural noise training, respectively. ``Int.'' denotes interactions.}
\label{tab:datasets_statistics}
\setlength{\tabcolsep}{1.2mm}{
\resizebox{0.48\textwidth}{!}{
\begin{tabular}{@{}llllll@{}}
\hline
& \multicolumn{1}{l}{\textbf{\#User}} & \multicolumn{1}{l}{\textbf{\#Item (C)}} & \multicolumn{1}{l}{\textbf{\#Int. (C)}} & \multicolumn{1}{l}{\textbf{\#Item (N)}} & \multicolumn{1}{l}{\textbf{\#Int. (N)}} \\ \hline
\textbf{Amazon-book} & 108,822 & 94,949 & 3,146,256 & 178,181 & 3,145,223 \\
\textbf{Yelp} & 54,574 & 34,395 & 1,402,736 & 77,405 & 1,471,675 \\
\textbf{ML-1M} & 5,949 & 2,810 & 571,531 & 3,494 & 618,297 \\ \hline
\end{tabular}
}}
\end{table}

We conduct experiments on three publicly available datasets in different scenarios. 1) \textbf{Amazon-book}\footnote{\url{https://jmcauley.ucsd.edu/data/amazon/.}} is from the Amazon review datasets,
which covers rich user interactions with extensive books. 2) \textbf{Yelp}\footnote{\url{https://www.yelp.com/dataset/.}} is a representative business dataset containing user reviews for different restaurants. 3) \textbf{ML-1M}\footnote{\url{https://grouplens.org/datasets/movielens/1m/.}} is a popular benchmark dataset with user ratings on movies. 

For all datasets, we first sort all interactions chronologically according to the timestamps. Thereafter, we consider three different training settings as follows. 1) \textbf{Clean training} discards user interactions with ratings $<$ 4, and then splits the sorted interactions into training, validation, and testing sets with the ratio of 7:2:1.
2) \textbf{Noisy training} keeps the same testing set of clean training, but adds some noisy interactions, including natural noises (\ie the interactions with ratings $<$ 4) and randomly sampled interactions into the training and validation sets. Note that we keep the numbers of noisy training and validation interactions on a similar scale as clean training for a fair comparison. 
3) \textbf{Temporal training}: to evaluate the effectiveness of temporal modeling, we additionally consider using timestamps for training, \ie modeling the user interaction sequences like sequential recommender models. The testing set is also the same as clean and noisy training for a fair comparison. 
The dataset statistics are summarized in Table~\ref{tab:datasets_statistics}.

\begin{table*}[t]
\setlength{\abovecaptionskip}{0.05cm}
\setlength{\belowcaptionskip}{0cm}
\caption{
Overall performance comparison between the baselines and DiffRec under clean training on three datasets. The best results are highlighted in bold and the second-best results are underlined. \% Improve. represents the relative improvements of DiffRec over the best baseline results. $*$ implies the improvements over the best baseline are statistically significant ($p$-value < 0.05) under one-sample t-tests.}
\label{tab:diffrec_clean}
\setlength{\tabcolsep}{2.0mm}{
\resizebox{\textwidth}{!}{
\begin{tabular}{l|cccc|cccc|cccc}
\hline
& \multicolumn{4}{c|}{\textbf{Amazon-book}} & \multicolumn{4}{c|}{\textbf{Yelp}} & \multicolumn{4}{c}{\textbf{ML-1M}} \\
\textbf{Methods} & \textbf{R@10} & \textbf{R@20}  & \textbf{N@10}  & \textbf{N@20}  & \textbf{R@10}  & \textbf{R@20}  & \textbf{N@10}  & \textbf{N@20}  & \textbf{R@10}  & \textbf{R@20}  & \textbf{N@10}  & \textbf{N@20}  \\ \hline
\textbf{MF} & 0.0437 & 0.0689 & 0.0264 & 0.0339 & 0.0341 & 0.0560 & 0.0210 & 0.0276 & 0.0876 & 0.1503 & 0.0749 & 0.0966 \\
\textbf{LightGCN} & 0.0534 & 0.0822 & 0.0325 & 0.0411 & 0.0540 & 0.0904 & 0.0325 & 0.0436 & 0.0987 & 0.1707 & 0.0833 & 0.1083 \\
\textbf{CDAE} & 0.0538 & 0.0737 & 0.0361 & 0.0422 & 0.0444 & 0.0703 & 0.0280 & 0.0360 & 0.0991 & 0.1705 & 0.0829 & 0.1078 \\
\textbf{MultiDAE} & 0.0571 & 0.0855 & 0.0357 & 0.0442 & 0.0522 & 0.0864 & 0.0316 & 0.0419 & 0.0995 & 0.1753 & 0.0803 & 0.1067 \\
\textbf{MultiDAE++} & 0.0580 & 0.0864 & 0.0363 & 0.0448 & 0.0544 & 0.0909 & 0.0328 & 0.0438 & {\ul 0.1009} & {\ul0.1771} & 0.0815 & 0.1079 \\
\textbf{MultiVAE} & {\ul 0.0628} & {\ul 0.0935} & {\ul 0.0393} & {\ul 0.0485} & {\ul 0.0567} & {\ul 0.0945} & {\ul 0.0344} & {\ul 0.0458} & 0.1007 & 0.1726 & 0.0825 & 0.1076 \\
\textbf{CODIGEM\footnotemark[6]} & 0.0300 & 0.0478 & 0.0192 & 0.0245 & 0.0470 & 0.0775 & 0.0292 & 0.0385 & 0.0972 & 0.1699 & {\ul 0.0837} & {\ul 0.1087} \\ \hline
\cellcolor{gray!16}\textbf{DiffRec} & \cellcolor{gray!16}\textbf{0.0695*} & \cellcolor{gray!16}\textbf{0.1010*} & \cellcolor{gray!16}\textbf{0.0451*} & \cellcolor{gray!16}\textbf{0.0547*} & \cellcolor{gray!16}\textbf{0.0581*} & \cellcolor{gray!16}\textbf{0.0960*} & \cellcolor{gray!16}\textbf{0.0363*} & \cellcolor{gray!16}\textbf{0.0478*} & \cellcolor{gray!16}\textbf{0.1058*} & \cellcolor{gray!16}\textbf{0.1787*} & \cellcolor{gray!16}\textbf{0.0901*} & \cellcolor{gray!16}\textbf{0.1148*} \\

\textbf{\% Improve.} & 10.67\% & 8.02\% & 14.76\% & 12.78\% & 2.47\% & 1.59\% & 5.52\% & 4.37\% & 4.86\% & 0.90\% & 9.21\% & 6.69\% \\ \hline
\end{tabular}
}}

\end{table*}

\subsubsection{\textbf{Baselines}}

\footnotetext[6]{The results on ML-1M differ from those reported in~\cite{walker2022recommendation}, owing to different data processing procedures. \cite{walker2022recommendation} did not sort and split the training/testing sets according to timestamps; however, temporal splitting aligns better with the real-world testing.}

We compare DiffRec with competitive baselines, including generative methods, and non-generative methods.

\begin{itemize}[leftmargin=*]
    \item \textbf{MF}~\cite{rendle2009bpr} is one of the most representative collaborative filtering methods based on matrix factorization. 
    
    \item \textbf{LightGCN}~\cite{he2020lightgcn} learns user and item representations via the linear neighborhood aggregation on graph convolution networks.
    
    \item \textbf{CDAE}~\cite{wu2016collaborative} trains an Auto-Encoder (AE) to recover the original user interactions from the randomly corrupted interactions.
    
    \item \textbf{MultiDAE}~\cite{liang2018variational} uses dropout to corrupt the interactions and recover them via an AE with the multinomial likelihood.

    \item \textbf{MultiDAE++} is designed by us by adding noises to
    corrupt interactions similar to DiffRec and training a MultiDAE to recover clean interactions in a single decoding step. The added noises in MultiDAE++ are the same as DiffRec while DiffRec learns to denoise little by little in the reverse process. 

    \item \textbf{MultiVAE}~\cite{liang2018variational} utilizes VAEs to model the interaction generation process, where the posterior is approximated by an encoder. 
    
    \item \textbf{CODIGEM}~\cite{walker2022recommendation} is a generative model using the diffusion process, which adopts multiple AEs to model the reverse generation yet only utilizes the first AE for interaction prediction. 
\end{itemize}

\noindent{\textbf{Evaluation}}.
We follow the full-ranking protocol~\cite{he2020lightgcn} by ranking all the non-interacted items for each user. For performance comparison, we adopt two widely used metrics Recall@$K$ (R@$K$) and NDCG@$K$ (N@$K$) over the top-$K$ items, where $K$ is set as 10 or 20.

\subsubsection{\textbf{Hyper-parameters Settings.}}

 We select the best hyper-parameters according to Recall@20 on the validation set. 
We tune the learning rates of all models in $\{1e^{-5}, 1e^{-4}, 1e^{-3}, 1e^{-2}\}$. 
As to model-specific hyper-parameters, the search scopes are as follows.

{- MF \& LightGCN.} The dropout ratio is selected from $\{0.1, 0.2,$ $0.3, 0.4, 0.5\}$. The weight decay is chosen from $\{1e^{-6}, 1e^{-5}, 1e^{-4}\}$ and the number of propagation layers is searched in $\{1, 2, 3\}$. 

{- CDAE \& MultiDAE \&  MultiDAE++ \& MultVAE.} We tune the weight decay and dropout ratio in the scopes of $\{0, 1e^{-3}, 1e^{-1}\}$ and $\{0.1, 0.3, 0.5\}$, respectively. 
Besides, we choose the activation function of CDAE from $\{\text{sigmoid}, \text{relu}, \text{tanh}\}$. 
As to MultVAE, the regularization strength $\beta$ and the annealing step are searched in $\{0, 0.3, 0.5, 0.7\}$ and $\{0, 200, 500\}$, respectively. The noises for MultiDAE++ are fixed consistently with DiffRec. 
The hidden size is set to the default value of $[200,600]$. 

{- CODIGEM.} The diffusion step is chosen from $\{2, 5, 10, 40, 50,\\100\}$ and the noise $\beta$ at each step is tuned in range of $\{5e^{-5}, 1e^{-4}, \\5e^{-4}\}$. The hidden sizes of the multiple five-layer AEs are set to the default value of 200.

{- DiffRec \& L-DiffRec \& T-DiffRec.} The step embedding size is fixed at 10. We choose the hidden size of the MLP of $p_\theta(\bm{x}_{t-1}|\bm{x}_{t})$ in $\{[300], [200,600], [1000]\}$. The diffusion step $T$ and the inference step $T'$ are tuned in $\{2, 5, 10, 40, 50, 100\}$ and $\{0, \frac{T}{4}, \frac{T}{2}\}$, respectively. Besides, the noise scale $s$, the noise lower bound $\alpha_{\text{min}}$, the noise upper bound $\alpha_{\text{max}}$ are searched in $\{0, 1e^{-5}, 1e^{-4}, 5e^{-3}, 1e^{-2}, 1e^{-1}, \\5e^{-1}\}$, $\{5e^{-4}, 1e^{-3}, 5e^{-3}\}$, and $\{5e^{-3}, 1e^{-2}, 2e^{-2}\}$, respectively. 
As to L-DiffRec, the dimension of $\bm{z}_0$ is set to 300 and the category number $C$ is chosen from $\{1,2,3,4,5\}$. 
For T-DiffRec, $w_{\text{min}}$ is tuned in $\{0.1, 0.3, 0.5\}$ and $w_{\text{max}}$ is set to $1$. More details are included in our released code.

All experiments are done using a single Tesla-V100 GPU, except for ACVAE in Table~\ref{tab:t-diffrec_clean} using A40 due to high computing costs.

\subsection{Analysis of DiffRec (RQ1)}
\label{sec:performance_diffrec}

\begin{table*}[t]
\setlength{\abovecaptionskip}{0.05cm}
\setlength{\belowcaptionskip}{0cm}
\caption{Performance comparison between DiffRec, the best generative baseline (MultiVAE), and the best non-generative baseline (LightGCN) under noisy training with natural noises. 
}
\label{tab:diffrec_noisy}
\setlength{\tabcolsep}{4mm}{
\resizebox{1\textwidth}{!}{
\begin{tabular}{l|cccc|cccc|cccc}
\hline
 & \multicolumn{4}{c|}{\textbf{Amazon-book}} & \multicolumn{4}{c|}{\textbf{Yelp}} & \multicolumn{4}{c}{\textbf{ML-1M}} \\ 
 & \textbf{R@10}  & \textbf{R@20}  & \textbf{N@10}  & \textbf{N@20}  & \textbf{R@10}  & \textbf{R@20}  & \textbf{N@10}  & \textbf{N@20}  & \textbf{R@10}  & \textbf{R@20}  & \textbf{N@10}  & \textbf{N@20}  \\ \hline
\textbf{LightGCN} & 0.0400 & 0.0659 & 0.0231 & 0.0308 & 0.0466 & 0.0803 & 0.0278 & 0.0379 & 0.0648 & 0.1226 & {\ul0.0470} & 0.0679 \\
\textbf{MultiVAE} & {\ul0.0536} & {\ul0.0820} & {\ul0.0316} & {\ul0.0401} & {\ul0.0494} & {\ul0.0834} & {\ul0.0293} & {\ul0.0396} & {\ul0.0653} & \textbf{0.1247} & 0.0469 & {\ul0.0680} \\
\textbf{DiffRec} & \textbf{0.0546} & \textbf{0.0822} & \textbf{0.0335} & \textbf{0.0419} & \textbf{0.0507} & \textbf{0.0853} & \textbf{0.0309} & \textbf{0.0414} & \textbf{0.0658} & {\ul0.1236} & \textbf{0.0488} & \textbf{0.0703} \\ \hline
\end{tabular}
}}
\end{table*}
\subsubsection{\textbf{Clean Training}}

We first present the comparison between DiffRec and the baselines under clean training without using timestamps in Table~\ref{tab:diffrec_clean}, from which we have the following observations.

\begin{itemize}[leftmargin=*]
    \item Most generative methods (\ie MultiVAE, MultiDAE, MultiDAE++, CDAE) usually yield better performance than MF and LightGCN. These superior results are possibly attributed to the alignment between the generative modeling and the real-world interaction generation procedure. 
    Among all generative methods, MultiVAE reveals impressive performance, especially on Amazon-book and Yelp. 
    This is because it utilizes variational inference and multinomial likelihood~\cite{liang2018variational}, leading to stronger generation modeling. 
    
    \item In all cases, our revised MultiDAE++ consistently outperforms MultiDAE. This implies the effectiveness of denoising training on enhancing the representation abilities of generative models. 
    Besides, CODIGEM performs worse compared to LightGCN and other generative methods. This is fair because although multiple AEs are trained to model the forward and reverse processes, CODIGEM only uses the first AE for inference, and thus it is essentially learning a MultiDAE with the noises at a small scale. The inferior performance of CODIGEM than MultiVAE is also consistent with the results in Table 2 of~\cite{walker2022recommendation}.

    \item DiffRec significantly achieves superior performance on three datasets. The large improvements over VAE-based methods validate the superiority of applying DMs for recommender systems. Such improvements result from that 1) DiffRec is capable of modeling complex distributions via gradually learning each denoising transition step from $t$ to $t-1$ with shared neural networks~\cite{rombach2022high}; 
    2) DiffRec utilizes simple forward corruption for tractable posterior distribution, alleviating the intrinsic trade-off between the tractability and representation ability of VAE-based methods; 
    and 3) notably, the scheduled noises for corruption in Eq. (\ref{eq:noise_schedule}) ensure personalized preference modeling (\cf Section~\ref{sec:method_discussion}).
    
\end{itemize}

\subsubsection{\textbf{Noisy Training}}

\begin{figure}[t]
\vspace{-0.3cm}
\setlength{\abovecaptionskip}{-0.3cm}
\setlength{\belowcaptionskip}{-0.10cm}
  \centering 
  \hspace{-0.3in}
  \subfigure{
    \includegraphics[width=1.58in]{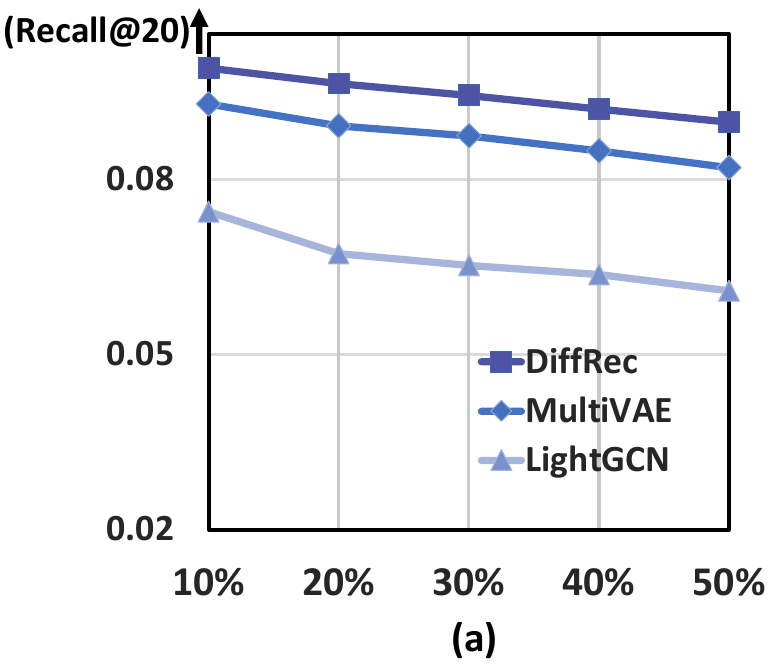}} 
  \hspace{-0.105in}
  \subfigure{
     \includegraphics[width=1.58in]{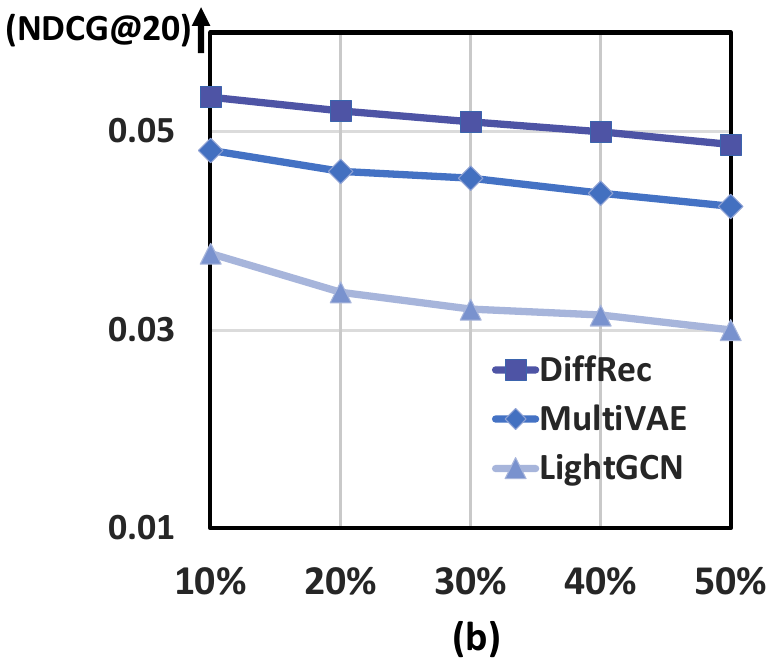}} 
\caption{Performance comparison of noisy training with random noises on Amazon-book.}
  \label{fig:diffrec_random_noise}
\end{figure}
In real-world recommender systems, collected user interactions in implicit feedback naturally contain false-positive and false-negative items. 
To analyze the performance of DiffRec on learning from noisy interactions, we compare DiffRec with the best non-generative method LightGCN and the best generative method MultiVAE under two noisy settings: 
1) \textbf{natural noises}, where we randomly add some false-positive interactions with ratings $<$ 4 as positive ones to the training and validation sets (see Section~\ref{sec:datasets}); 
and 2) \textbf{random noises}, where we randomly add a proportion of non-interacted items as positive interactions for each user. We summarize the performance of natural noises in Table~\ref{tab:diffrec_noisy} and the results of random noises with the noise proportion ranging from 10\% to 50\% in Figure~\ref{fig:diffrec_random_noise}. 
In Figure~\ref{fig:diffrec_random_noise}, we only show the results on Amazon-book to save space as we have similar observations on Yelp and ML-1M.

From Table~\ref{tab:diffrec_noisy}, we can observe that DiffRec usually surpasses MultiVAE and LightGCN, verifying the strong robustness of DiffRec against natural noises. This is reasonable since such false-positive interactions are essentially corrupted interactions and DiffRec is intrinsically optimized to recover clean interactions iteratively from the corruption. 
By contrast, LightGCN is vulnerable to noisy interactions because it might amplify the negative effect of noises by emphasizing high-order propagation, thus leading to poor performance. In addition, the comparable results on ML-1M are because this dense dataset is relatively easier for prediction.

From the results in Figure~\ref{fig:diffrec_random_noise}, we can find: 1) from adding 10\% to 50\% random noises, the performance of LightGCN, MultiVAE, and DiffRec gradually declines. This observation makes sense because it is harder to predict user preference as noises increase. 
Nevertheless, 2) DiffRec still outperforms MultiVAE and LightGCN even under a large scale of noises. The reason is that DiffRec is trained under different noise scales at each step, facilitating the recovery of real interactions from heavily corrupted interactions.

\begin{figure}[t]
\vspace{-0.3cm}
\setlength{\abovecaptionskip}{-0.10cm}
\setlength{\belowcaptionskip}{0cm}
  \centering 
  \subfigure{
    \includegraphics[width=1.38in]{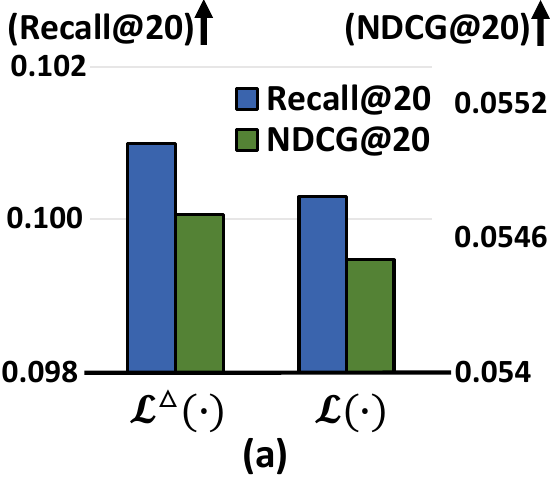}} 
  \hspace{-0.02in}
  \subfigure{
    \includegraphics[width=1.72in]{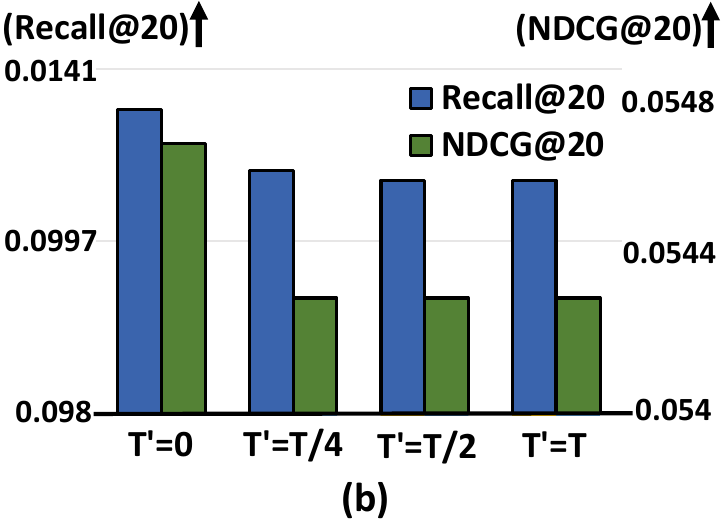}} 
\caption{Effects of $\mathcal{L}^\bigtriangleup(\cdot)$, $\mathcal{L}(\cdot)$, and $T'$, where $\mathcal{L}^\bigtriangleup(\cdot)$ and $\mathcal{L}(\cdot)$ mean importance sampling in Eq. (\ref{eq:L_importance}) and uniform sampling in Eq. (\ref{eq:L}), respectively. $T'$ is the inference step.}
  \label{fig:different_design}
  \vspace{-0.40cm}
\end{figure}

\begin{figure}[t]
\setlength{\abovecaptionskip}{-0.10cm}
\setlength{\belowcaptionskip}{0cm}
  \centering 
  \hspace{-0.2in}
  \subfigure{
    \includegraphics[width=1.6in]{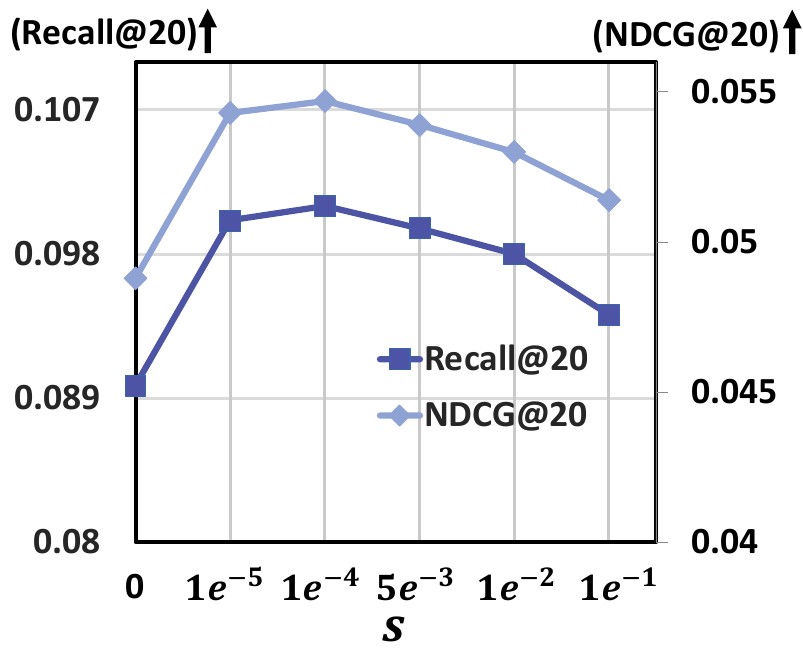}} 
  \hspace{-0.02in}
  \subfigure{
     \includegraphics[width=1.6in]{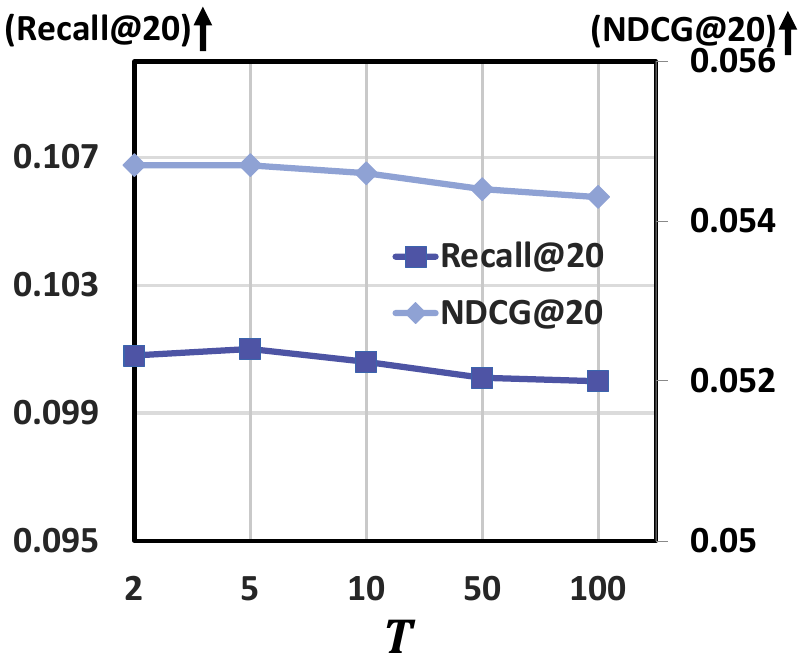}} 
\caption{Effects of the noise scale $s$ and diffusion step $T$.}
  \label{fig:hp}
\end{figure}

\subsubsection{\textbf{In-depth Analysis}}\label{sec:in_depth}
We further explore the effects of different designs in DiffRec such as importance sampling, $\bm{x}_0$-ELBO, inference step $T'$, and noise scales. 
The results on Amazon-book are reported in Figure~\ref{fig:different_design} while the results on Yelp and ML-1M with similar observations are omitted to save space.

\begin{table*}[t]
\setlength{\abovecaptionskip}{0.05cm}
\setlength{\belowcaptionskip}{0cm}
\caption{Performance comparison between L-DiffRec and DiffRec under natural noise training on three datasets.}
\label{tab:l-diffrec_noisy}
\setlength{\tabcolsep}{0.06mm}{
\resizebox{\textwidth}{!}{
\begin{tabular}{l|cccc|cccc|cccc}
\hline
 & \multicolumn{4}{c|}{\textbf{Amazon-book}} & \multicolumn{4}{c|}{\textbf{Yelp}} & \multicolumn{4}{c}{\textbf{ML-1M}} \\
 & \textbf{R@10}  & \textbf{R@20}  & \textbf{N@10}  & \textbf{N@20}  & \textbf{R@10}  & \textbf{R@20}  & \textbf{N@10}  & \textbf{N@20}  & \textbf{R@10}  & \textbf{R@20}  & \textbf{N@10}  & \textbf{N@20}  \\ \hline
\textbf{DiffRec} & 0.0546 & 0.0822 & 0.0335 & 0.0419 & 0.0507 & 0.0853 & 0.0309 & 0.0414 & 0.0658 & 0.1236 & 0.0488 & 0.0703 \\
\textbf{L-DiffRec} & \textcolor{white}{666}0.0586$^{\imp{+7.3\%}}$ & \textcolor{white}{666}0.0876$^{\imp{+6.6\%}}$ & \textcolor{white}{666}0.0347$^{\imp{+3.6\%}}$ & \textcolor{white}{666}0.0434$^{\imp{+3.6\%}}$ & \textcolor{white}{666}0.0521$^{\imp{+2.8\%}}$ & \textcolor{white}{666}0.0876$^{\imp{+2.7\%}}$ & \textcolor{white}{666}0.0311$^{\imp{+0.7\%}}$ & \textcolor{white}{666}0.0419$^{\imp{+1.2\%}}$ & \textcolor{white}{666}0.0665$^{\imp{+1.1\%}}$ & \textcolor{white}{666}0.1272$^{\imp{+2.9\%}}$ & \textcolor{white}{666}0.0493$^{\imp{+1.0\%}}$ & \textcolor{white}{666}0.0710$^{\imp{+1.0\%}}$ \\ \hline
\end{tabular}
}}
\vspace{-0.3cm}
\end{table*}

\begin{table}[t]
\setlength{\abovecaptionskip}{0.05cm}
\setlength{\belowcaptionskip}{0cm}
\caption{Performance of $\bm{\epsilon}$-ELBO on ML-1M.}
\label{tab:indepth_mult_epsilon}
\setlength{\tabcolsep}{4.0mm}{
\resizebox{0.48\textwidth}{!}{
\begin{tabular}{lllll}
\hline
\textbf{Variants} & \multicolumn{1}{c}{\textbf{R@10} } & \multicolumn{1}{c}{\textbf{R@20} } & \multicolumn{1}{c}{\textbf{N@10} } & \multicolumn{1}{c}{\textbf{N@20} } \\ \hline
\textbf{DiffRec ($\bm{x}_0$-ELBO)} & 0.1058 & 0.1787 & 0.0901 & 0.1148 \\
\textbf{$\bm{\epsilon}$-ELBO} & 0.0157 & 0.0266 & 0.0170 & 0.0204 \\ \hline
\end{tabular}
}}
\end{table}

\vspace{3pt}
\noindent{\textbf{$\bullet$ Importance sampling.} 
We compare the performance between importance sampling ($\mathcal{L}^\bigtriangleup(\cdot)$ in Eq. (\ref{eq:L_importance})) and uniform sampling ($\mathcal{L}(\cdot)$ in Eq. (\ref{eq:L})) in Figure~\ref{fig:different_design}(a). 
The declined performance of $\mathcal{L}(\cdot)$ validates the effectiveness of importance sampling, which assigns large sampling probabilities to the large-loss steps and thus focuses on ``hard'' denoising steps for optimization.

\vspace{3pt}
\noindent{\textbf{$\bullet$ Effect of inference step $T'$}}.
We vary $T'$ from $0$ to $T$ during inference and show the results in Figure~\ref{fig:different_design}(b), from which we can find that using $T'=0$ achieves better performance. It makes sense because collected interactions from real-world data naturally contain noises, and too much corruption might hurt personalization. 
In addition, the results are comparable when $T'$ changes from $T/4$ to $T$, possibly because the scheduled noise scale is relatively small, leading to minor changes in the ranking positions of top-$K$ items.

\begin{figure*}[t]
\setlength{\abovecaptionskip}{-0.10cm}
\setlength{\belowcaptionskip}{0cm}
  \centering 
  \hspace{-0.7in}
  \subfigure{
    \includegraphics[width=1.67in]{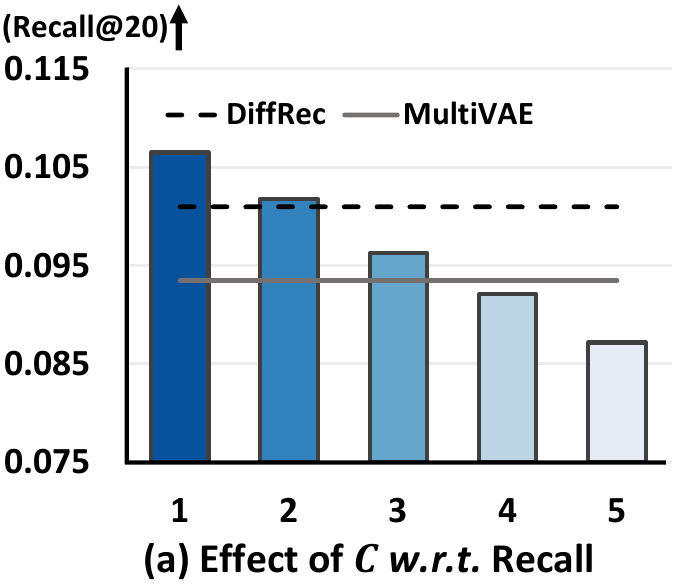}} 
  \hspace{-0.105in}
  \subfigure{
     \includegraphics[width=1.67in]{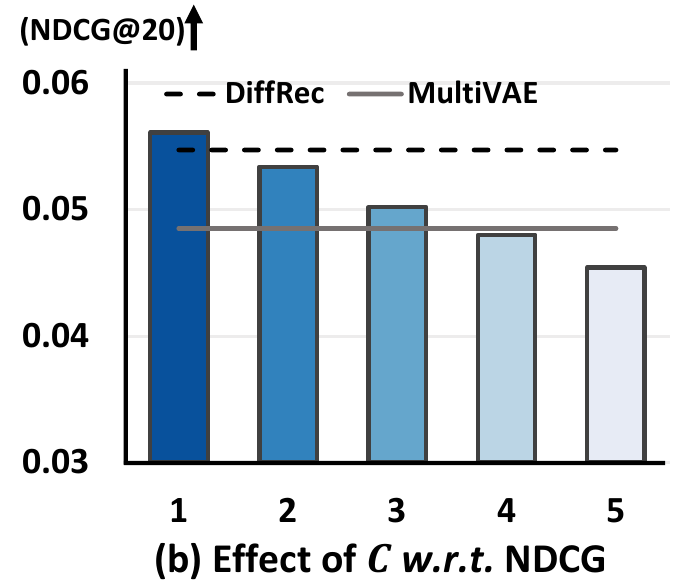}} 
\hspace{-0.105in}
  \subfigure{
    \includegraphics[width=1.67in]{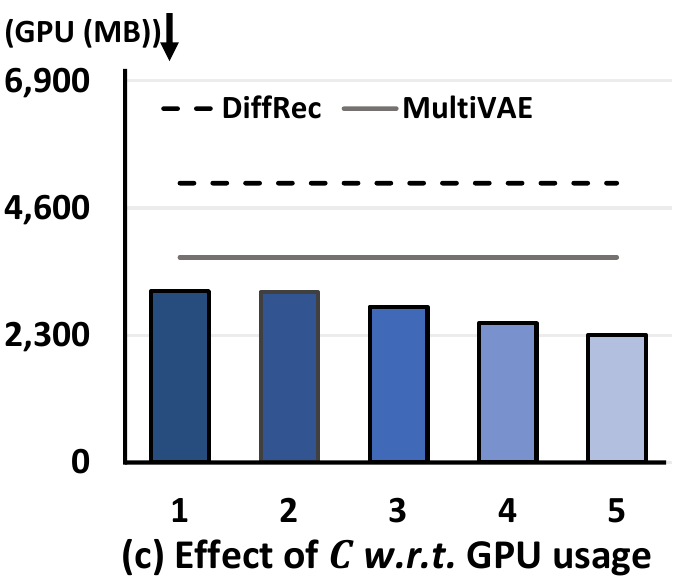}} 
\hspace{-0.105in}
  \subfigure{
    \includegraphics[width=1.67in]{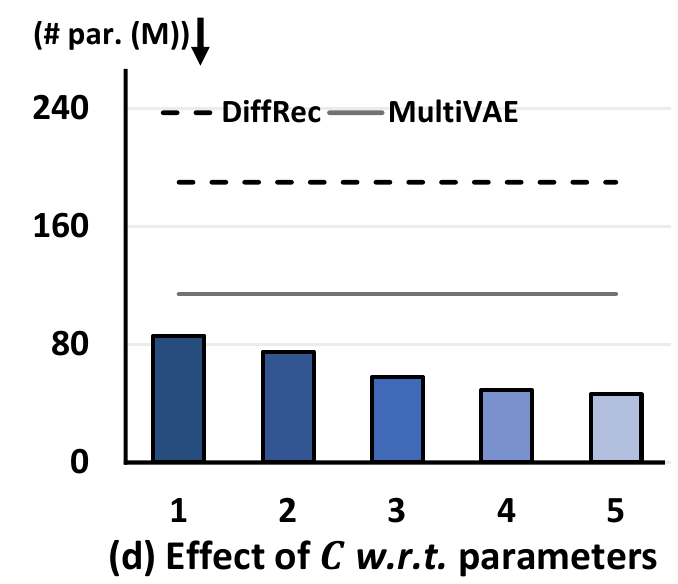}} 
  \hspace{-0.7in} 
\caption{Effects of the clustering category number of L-DiffRec on Amazon-book under clean training.}
  \label{fig:ldiffrec_category}
  \vspace{-0.20cm}
\end{figure*}

\vspace{3pt}
\noindent{\textbf{$\bullet$ Effects of noise scale $s$ and step $T$.}}
In DiffRec, there are two important hyper-parameters: diffusion step $T$ and noise scale $s$. To analyze their effect, we vary $s$ at different scales and change $T$ from 2 to 100, respectively. From the results in Figure~\ref{fig:hp}, we can observe that: 
1) as the noise scale increases, the performance first rises compared to training without noise ($s=0$), verifying the effectiveness of denoising training. However, enlarging noise scales degrades the performance due to corrupting the personalization. Hence, we should carefully choose a relatively small noise scale (\eg $1e^{-4}$) as discussed in Section~\ref{sec:method_discussion}. 
And 2) the performance fluctuation \wrt $T$ indicates that increasing diffusion steps has little effects on accuracy due to the relatively small noises in the forward process. Considering $T$ being too large will cause high computing burdens, we choose $T=5$ for good performance as well as low costs. 

\vspace{3pt}
\noindent{\textbf{$\bullet$ $\bm{x}_0$-ELBO} vs. \textbf{$\bm{\epsilon}$-ELBO}}. 
The comparison between predicting $\bm{x}_0$ and $\bm{\epsilon}$ ($\bm{\epsilon}$-ELBO, introduced in Section~\ref{sec:method_discussion}) on ML-1M is in Table~\ref{tab:indepth_mult_epsilon}. The results of $\bm{\epsilon}$-ELBO on Amazon-book and Yelp are close to zero due to severer data sparsity, and thus are omitted to save space. We attribute the worse results of $\bm{\epsilon}$-ELBO to the difficulty of predicting randomly sampled noises via an MLP. Besides, the reduced noise scales $s$ may also enhance the prediction difficulty because the noises of different steps are becoming small with minor differences. We leave the further theoretical analysis to future work.

\vspace{-0.3cm}
\subsection{Analysis of L-DiffRec (RQ2)}\label{sec:performance_l-diffrec}
To analyze L-DiffRec performance \wrt accuracy and resource costs, we evaluate L-DiffRec on three datasets under clean and noisy training. Moreover, we examine the effect of clustering category numbers to facilitate the future application of L-DiffRec. 

\subsubsection{\textbf{Clean Training.}}
From Table~\ref{tab:l-diffrec_clean}, we can find that L-DiffRec significantly outperforms MultiVAE with fewer resource costs (38.39\% parameters and 10.61\% GPU memory reduced on average), justifying the superiority of L-DiffRec. 
Meanwhile, it drastically lowers the costs of DiffRec with comparable accuracy, \ie reducing 56.17\% parameters and 24.64\% GPU usage on average. The comparable accuracy might be attributed to that the diffusion in the interaction space has redundant information and the dimension compression via clustering does not lose key information. 
The remarkable decline of resources is due to that 1) item clustering reduces the parameters of the encoder and decoder; 2) the latent diffusion lessens the parameters of the denoising MLP with $\theta$.
With significantly fewer resource costs, L-DiffRec has the potential to enable large-scale item prediction in industrial scenarios. 

\begin{table}[t]
\setlength{\abovecaptionskip}{0.05cm}
\setlength{\belowcaptionskip}{0cm}
\caption{Performance of L-DiffRec with $C=2$, DiffRec, and MultiVAE under clean training. ``par." denotes parameters.}
\label{tab:l-diffrec_clean}
\setlength{\tabcolsep}{0.3mm}{
\resizebox{0.48\textwidth}{!}{
\begin{tabular}{@{}llcccccc@{}}
\toprule
\textbf{Datasets} & \textbf{Method} & \textbf{R@10}$\uparrow$ & \textbf{R@20}$\uparrow$ & \textbf{N@10}$\uparrow$ & \textbf{N@20}$\uparrow$ & \textbf{\#par.(M)}$\downarrow$ & \textbf{GPU(MB)}$\downarrow$ \\ \hline 
\multirow{3}{*}{\begin{tabular}[c]{@{}l@{}}\textbf{Amazon}\\ \textbf{-book}\end{tabular}} & \textbf{MultiVAE} & 0.0628 & 0.0935 & 0.0393 & 0.0485 & {\ul 114} & {\ul3,711}  \\
 & \textbf{DiffRec} & \textbf{0.0695} & {\ul0.1010} & \textbf{0.0451} & \textbf{0.0547} & 190 & 5,049  \\
 & \textbf{L-DiffRec} & {\ul0.0694} & \textbf{0.1028} & {\ul0.0440} & {\ul0.0540} & \textbf{75} & \textbf{3,077}  \\ \hline
 
\multirow{3}{*}{\textbf{Yelp}} & \textbf{MultiVAE} & 0.0567 & 0.0945 & 0.0344 & 0.0458 & {\ul42} & {\ul1,615}  \\
 & \textbf{DiffRec} & {\ul0.0581} & {\ul0.0960} & \textbf{0.0363} & \textbf{0.0478} & 69 & 2,103  \\
 & \textbf{L-DiffRec} & \textbf{0.0585} & \textbf{0.0970} & {\ul0.0353} & {\ul0.0469}  & \textbf{29} & \textbf{1,429}\\ \hline
 
\multirow{3}{*}{\textbf{ML-1M}} & \textbf{MultiVAE} & 0.1007 & 0.1726 & 0.0825 & 0.1076  & 4 & 497\\
 & \textbf{DiffRec} & {\ul0.1058} & {\ul0.1787} & \textbf{0.0901} & \textbf{0.1148}  & {\ul4} & {\ul495}\\
 & \textbf{L-DiffRec} & \textbf{0.1060} & \textbf{0.1809} & {\ul0.0868} & {\ul0.1122}  & \textbf{2} & \textbf{481}\\ \bottomrule
\end{tabular}
}}
\end{table}

\subsubsection{\textbf{Noisy Training.}}
The resource costs of noisy training are the same as clean training while we observe that 
L-DiffRec consistently outperforms DiffRec under noisy training as shown in Table~\ref{tab:l-diffrec_noisy}. One possible reason is that some clustered categories have few interactions, which are more likely to be false-positive interactions. The effect of such noises is weakened after representation compression via item clustering. 

\begin{table*}[t]
\setlength{\abovecaptionskip}{0.02cm}
\setlength{\belowcaptionskip}{0cm}
\caption{Performance comparison between DiffRec variants and a SOTA sequential baseline ACVAE. The models are trained using timestamps. The results on ML-1M are similar to Amazon-Book and omitted to save space. ``par." denotes parameters.}
\label{tab:t-diffrec_clean}
\setlength{\tabcolsep}{2mm}{
\resizebox{1\textwidth}{!}{
\begin{tabular}{l|cccccc|cccccc}
\hline
 & \multicolumn{6}{c|}{\textbf{Amazon-book}} & \multicolumn{6}{c}{\textbf{Yelp}} \\
 & \textbf{R@10}$\uparrow$ & \textbf{R@20}$\uparrow$ & \textbf{N@10}$\uparrow$ & \textbf{N@20}$\uparrow$ & \textbf{\#par. (M)}$\downarrow$ & \textbf{GPU (MB)}$\downarrow$  & \textbf{R@10}$\uparrow$ & \textbf{R@20}$\uparrow$ & \textbf{N@10}$\uparrow$ & \textbf{N@20}$\uparrow$ & \textbf{\#par. (M)}$\downarrow$ & \textbf{GPU (MB)}$\downarrow$ \\ \hline
\textbf{ACVAE} & 0.0770 & 0.1107 & 0.0547 & 0.0647  & 13 & 37,711 & 0.0567 & 0.0947 & 0.0342 & 0.0456  & 5 & 14,697\\ \hline
\textbf{DiffRec} & 0.0695 & 0.1010 & 0.0451 & 0.0547 & 190 & 5,049  & 0.0581 & 0.0960 & 0.0363 & 0.0478  & 69 & 2,107\\
\cellcolor{gray!16}\textbf{T-DiffRec} & \cellcolor{gray!16}0.0819 & \cellcolor{gray!16}0.1139 & \cellcolor{gray!16}0.0565 & \cellcolor{gray!16}0.0661 & \cellcolor{gray!16}190 & \cellcolor{gray!16}5,049  & \cellcolor{gray!16}0.0601 & \cellcolor{gray!16}0.0987 & \cellcolor{gray!16}0.0377 & \cellcolor{gray!16}0.0494  & \cellcolor{gray!16}69 & \cellcolor{gray!16}2,107\\ \hline

\textbf{L-DiffRec} & 0.0694 & 0.1028 & 0.0440  & 0.0540 & 75 & 3,077  & 0.0585 & 0.0970 & 0.0353 & 0.0469  & 29 & 1,429\\
\cellcolor{gray!16}\textbf{LT-DiffRec} & \cellcolor{gray!16}0.0838 & \cellcolor{gray!16}0.1188 & \cellcolor{gray!16}0.0560 & \cellcolor{gray!16}0.0665 & \cellcolor{gray!16}75 & \cellcolor{gray!16}3,077  & \cellcolor{gray!16}0.0604 & \cellcolor{gray!16}0.0982 & \cellcolor{gray!16}0.0369 & \cellcolor{gray!16}0.0484  & \cellcolor{gray!16}29 & \cellcolor{gray!16}1,429\\ \hline
\end{tabular}
}}
\vspace{-0.2cm}
\end{table*}

\subsubsection{\textbf{Effect of category number.}}
To inspect the effect of category number on L-DiffRec, we compare the results with clustering category numbers changing from 1 to 5 on Amazon-book. We omitted similar results on Yelp and ML-1M to save space. From Figure~\ref{fig:ldiffrec_category}, we can find that:
1) the Recall, NDCG, GPU usage, and parameters decline as the category number $C$ increases as shown in Figure~\ref{fig:ldiffrec_category}(a) and (b). This is reasonable since increasing $C$ will reduce the parameters, hurting the representation ability.
2) The resource costs are substantially reduced compared to DiffRec and MultiVAE even if clustering is disabled ($C=1$). This is due to the significant parameter reduction of the denoising MLP via latent diffusion. 
And 3) L-DiffRec is comparable with DiffRec when $C=1$ or $2$ while L-DiffRec consistently outperforms MultiVAE when $C=1, 2,$ or $3$. As such, L-DiffRec can save extensive resources with comparable or superior accuracy by carefully choosing $C$.

\vspace{-0.3cm}
\subsection{Analysis of T-DiffRec (RQ3)}\label{sec:performance_t-diffrec}
To verify the effectiveness of T-DiffRec on temporal modeling, we compare T-DiffRec and LT-DiffRec with a SOTA sequential recommender model ACVAE~\cite{xie2021adversarial}, which employs VAE with contrastive learning and adversarial training for recommendation.

From Table~\ref{tab:t-diffrec_clean}, we have the following observations:
1) T-DiffRec and LT-DiffRec perform better than DiffRec and L-DiffRec by a large margin, justifying the effectiveness of the proposed time-aware reweighting strategy on temporal modeling;
2) the superior performance of T-DiffRec and LT-DiffRec than ACVAE is attributed to both capturing temporal shifts and conducting diffusion processes, leading to more accurate and robust user representations;  
3) despite more parameters, DiffRec-based methods consume much less GPU memory than ACVAE, thus reducing computing costs; 
4) it is highlighted that LT-DiffRec yields comparable performance to T-DiffRec with fewer parameters, which is consistent with observations in Section~\ref{sec:performance_l-diffrec};
and 5) the relatively small improvements of T-DiffRec over DiffRec on Yelp and the inferior results of ACVAE than DiffRec on Yelp are because user preference over food is relatively stable and the temporal shifts are limited. As such, considering temporal information receives minor benefits.

\vspace{-0.2cm}
\section{Related Work}
\label{sec:related_work}

\textbf{$\bullet$ Generative recommendation.}
Discriminative recommender models~\cite{Liu2021IMPGCN,wei2022causal} usually predict user-item interaction probabilities given the user and item representations.
Although discriminative methods are cost-friendly, generative models can better learn collaborative signals between items due to simultaneously modeling the predictions over all items~\cite{yu2019vaegan,ren2020sequential}.
Besides, generative models are specialized to capture the complex and non-linear relations between user preference and interactions as detailed in~\cite{li2015deep,shenbin2020recvae,li2017collaborative}. 
Existing generative recommender models can be roughly divided into two groups: GAN-based methods~\cite{guo2020ipgan,xu2022negative} and VAE-based methods~\cite{ma2019learning,liu2021variation}. 
GAN-based approaches utilize adversarial training~\cite{wang2017irgan,he2018adversarial,wang2022generative,wu2019pd} to optimize the generator for predicting user interactions~\cite{chen2022generative,gao2021recommender,jin2020sampling}. 
As to VAE-based methods~\cite{zhang2017autosvd++,ma2019learning}, they mainly learn an encoder for posterior estimation~\cite{higgins2017beta,ren2022variational}, and a decoder to predict the interaction probabilities over all items~\cite{wang2022causal}. For example, the most representative MultiVAE~\cite{liang2018variational} achieves impressive performance by variational modeling. 

Despite their success, DMs have shown great advantages over GANs and VAEs such as low instability and high generation quality in diverse tasks, including image synthesis~\cite{song2020denoising}, text generation~\cite{hoogeboom2021argmax}, and audio generation~\cite{huang2022prodiff}. As such, we consider revising DMs for generative recommendation. 

\vspace{3pt}
\noindent\textbf{$\bullet$ Diffusion models.}
DMs recently have shown the capability of high-quality generation~\cite{croitoru2022diffusion, popov2021grad}, covering conditional generation~\cite{chao2022denoising, liu2023more,ho2020denoising,rombach2022high} and unconditional generation~\cite{austin2021structured,lam2021bilateral}.

In spite of their success, utilizing DMs for recommendation receives little scrutiny. CODIGEM~\cite{walker2022recommendation} claims to generate recommendation via DMs, which however is essentially a noise-based MultiDAE method~\cite{liang2018variational} with inferior performance (\cf Table 2 in~\cite{walker2022recommendation}). 
Specifically, CODIGEM iteratively introduces noises step by step and utilizes multiple different AEs for the prediction at each step. During inference, it estimates the interaction probabilities merely using the first AE, and thus the remaining AEs are totally useless. As such, CODIGEM differs from our DiffRec that employs a shared MLP for the multi-step prediction and considers the multi-step denoising for inference. 
In addition, some studies on social recommendation consider information diffusion on social networks~\cite{WuLSHGW22,wu2019neural}. However, they mainly focus on the influence of social connections on user preference through diffusing processes~\cite{rafailidis2017recommendation}, which intrinsically differ from DiffRec.

\section{Conclusion and Future Work}
\label{sec:conclusion}

In this work, we proposed a novel DiffRec, which is a totally new recommender paradigm for generative recommender models. 
To ensure personalized recommendations, we reduced the noise scales and inference steps to corrupt users' interactions in the forward process. 
We also extended traditional DMs via two extensions to reduce the resource costs for large-scale item prediction and enable the temporal modeling of interaction sequences. 
Specifically, L-DiffRec clusters items for dimension compression and conducts diffusion processes in the latent space. Besides, T-DiffRec utilizes a time-aware reweighting strategy to capture the temporal patterns in users' interactions. 
Empirical results on three datasets under various settings validate the superiority of DiffRec with two extensions in terms of accuracy and resource costs.

This work opens up a new research direction for generative recommender models by employing DMs. 
Following this direction, many promising ideas deserve further exploration: 
1) although L-DiffRec and T-DiffRec are simple yet effective, it is beneficial to devise better strategies to achieve better model compression and encode temporal information (\eg transformers); 
2) it is meaningful to explore controllable or conditional recommendations based on DiffRec, \eg guiding the interaction prediction via a pre-trained classifier; 
and 3) exploring the effectiveness of more prior assumptions (\eg different noise assumptions other than Gaussian distribution) and diverse model structures is interesting. 

\clearpage

{
\tiny
\bibliographystyle{ACM-Reference-Format}
\balance
\bibliography{bibfile}
}


\end{document}